# Correction of local-linear elasticity for nonlocal residuals:
# Application to Euler-Bernoulli beams

## Mohamed Shaat*

*Engineering and Manufacturing Technologies Department, DACC, New Mexico State University, Las Cruces, NM 88003, USA*

*Mechanical Engineering Department, Zagazig University, Zagazig 44511, Egypt*

### Abstract

Complications exist when solving the field equation in the nonlocal field. This has been attributed to the complexity of deriving explicit forms of the nonlocal boundary conditions. Thus, the paradoxes in the existing solutions of the nonlocal field equation have been revealed in recent studies.

In the present study, a new methodology is proposed to easily determine the elastic nonlocal fields from their local counterparts without solving the field equation. This methodology depends on the iterative-nonlocal residual approach in which the sum of the nonlocal fields is treaded as a residual field. Thus, in this study the corrections of the local-linear elastic fields for the nonlocal residuals in materials are presented. These corrections are formed based on the general nonlocal theory. In the context of the general nonlocal theory, two distinct nonlocal parameters are introduced to form the constitutive equations of isotropic-linear elastic continua. In this study, it is demonstrated that the general nonlocal theory outperforms Eringen's nonlocal theory in accounting for the impacts of the material's Poisson's ratio on its mechanics. To demonstrate the effectiveness of the proposed approach, the corrections of the local static bending, vibration, and buckling characteristics of Euler-Bernoulli beams are derived. Via these corrections, bending, vibration, and buckling behaviors of simple-supported nonlocal Euler-Bernoulli beams are determined without solving the beam's equation of motion.

## 1. Introduction

Eringen's nonlocal theory has been proposed as a continuum theory that can model the long-range interatomic interactions of crystals [1-7]. In the context of this nonlocal theory, the nonlocal field is considered as the sum of the local field and a nonlocal residual [1-3]. An integral operator was proposed to incorporate these nonlocal residuals in the framework of linear elasticity [4-7]. Then, the integral operator

---------------------------------------------------

*Corresponding author: *E-mail address:* shaat@nmsu.edu; shaatscience@yahoo.com (M. Shaat).
Tel.: +15756215929

was replaced by a differential operator [8]. These two distinct forms of Eringen's nonlocal theory have been utilized in various studies. The nonlocal theory that depends on the integral operator, *i.e.* integral nonlocal theory, was used to investigate the dispersions of plane waves [9], stress concentrations near a crack tip [4], plasticity and damage of materials [10], softening plasticity [11], mechanics of nonlocal bars [12], and mechanics of nonlocal beams [13]. As for the differential nonlocal theory which depends on a differential operator, an uncountable list of studies were conducted; examples include [14-18].

Recently, Shaat [19, 20] pointed out that Eringen's nonlocal theory is inappropriate to model the mechanics of some materials including Si, Au, and Pt. Thus, Eringen's nonlocal theory cannot simultaneously fit the longitudinal and transverse acoustic dispersion curves of these materials. This has been attributed to the fact that Eringen's nonlocal theory forms the constitutive equations assuming only one nonlocal parameter for all material coefficients. This motivated Shaat [19, 20] to propose the general nonlocal theory. This general nonlocal theory introduces a distinct nonlocal parameter for each one of the material coefficients. Thus, for isotropic-linear elasticity, the general nonlocal theory forms the constitutive equations depending on two distinct nonlocal parameters, $\epsilon_1$ and $\epsilon_2$, as follows [19, 20]:

$$(1 - \epsilon_1 \boldsymbol{\nabla}^2)(1 - \epsilon_2 \boldsymbol{\nabla}^2) t_{ij} = \lambda (1 - \epsilon_2 \boldsymbol{\nabla}^2) \varepsilon_{rr} \delta_{ij} + 2\mu (1 - \epsilon_1 \boldsymbol{\nabla}^2) \varepsilon_{ij} \tag{1}$$

where $t_{ij}$ is the nonlocal stress field and $\varepsilon_{ij} = \frac{1}{2}(u_{i,j} + u_{j,i})$ is the infinitesimal strain field. $\lambda$ and $\mu$ are the material's Lame constants. $\boldsymbol{\nabla}^2$ denotes the Laplacian operator. In addition, the field equation can be written in the framework of the general nonlocal theory as follows:

$$\lambda (1 - \epsilon_2 \boldsymbol{\nabla}^2) u_{r,rj} \delta_{ij} + \mu (1 - \epsilon_1 \boldsymbol{\nabla}^2)(u_{i,jj} + u_{j,ij}) = (1 - \epsilon_1 \boldsymbol{\nabla}^2)(1 - \epsilon_2 \boldsymbol{\nabla}^2)(\rho \ddot{u}_i - f_i) \tag{2}$$

Simply, the general nonlocal theory recovers Eringen's nonlocal theory when $\epsilon_1 = \epsilon_2$. It was demonstrated in [20] that the general nonlocal theory allows for the simultaneous fitting of the longitudinal and transverse acoustic dispersions of different materials. This gives the general nonlocal theory the credits over Eringen's nonlocal theory.

In fact, solving the field equation (equation (2)) in the nonlocal fields is a challenging task for many reasons. First of all, for this field equation, higher-order essential boundary conditions are needed to represent the deformation of the boundary surface. However, because the nonlocal theory represents the continuum consisting of an infinite number of mass points, only the rigid motion and rotation of the boundary surface can be introduced as essential boundary conditions. Second, some studies revealed the paradoxes in the existing solutions of the differential nonlocal theory [21-26]. These paradoxes were attributed to the nonlocal boundary conditions. It was demonstrated that forming the natural boundary conditions of the differential nonlocal elasticity is a challenging task [15, 27].

The previously mentioned complications of solving the nonlocal field equation in the nonlocal fields are the motivations behind the present study. In this study, a new methodology that depends on the iterative-nonlocal residual approach [27] is proposed to easily determine the elastic nonlocal fields from their local counterparts without solving the complicated field equation. Thus, this study presents corrections of the local elastic fields for the nonlocal residuals in materials. The iterative-nonlocal residual approach was first proposed by Shaat [27] for determining the static bending of Kirchhoff plates. The merit of Shaat's iterative-nonlocal residual approach is that, instead of solving the nonlocal field equation in the nonlocal field, the field equation is solved in the local field where the local boundary conditions are the ones to be applied. This helps in resolving the complications associated with the nonlocal boundary conditions and the paradoxes in the existing solutions of the differential nonlocal theory. In the context of the iterative-nonlocal residual approach, a nonlocal residual field is formed iteratively and utilized to correct the local fields.

In this study, the iterative-nonlocal residual approach is utilized to correct the static bending, vibration, and buckling characteristics of local Euler-Bernoulli beams for the nonlocal residuals. The nonlocal residuals are formed based on the general nonlocal theory. To this end, the paper is organized as follows: first, correction factors for the static bending, free vibration natural frequencies, pre-buckling natural frequencies, and buckling and postbuckling characteristics of local Euler-Bernoulli beams for the nonlocal residuals are derived in section 2. Then, these correction factors are utilized to determine the aforementioned characteristics for nonlocal Euler-Bernoulli beams directly from their counterparts of local beams. In section 3, the proposed correction factors are obtained for simple supported nonlocal Euler-Bernoulli beams and utilized to determine their bending, vibration, and buckling characteristics. A part of the present study is devoted to demonstrate that the general nonlocal theory outperforms Eringen's nonlocal theory. Thus, when compared with Eringen's nonlocal theory, it is revealed that the general nonlocal theory (which considers two distinct nonlocal parameters) accounts for the impacts of the beam's Poisson's ratio on its bending, vibration, and buckling characteristics. Thus, a parametric study on the impacts of the nonlocal parameters and the beam's Poisson's ratio is presented. This parametric study demonstrate that, depending on the nonlocal parameters, Eringen's nonlocal theory may result in an inappropriate modeling of the nonlocal residual effects on Euler-Bernoulli beams.

## 2. Corrections of local elastic fields of Euler-Bernoulli beams for nonlocal residuals

A novel methodology to correct the local elastic fields, *i.e.* stress, strain, and displacement, for the nonlocal residuals in materials is proposed. Thus, instead of deriving solutions for the nonlocal elastic fields (which may be challenging for some nonlocal field problems), the local elastic fields are corrected to obtain the nonlocal elastic fields via correction factors. These correction factors are derived based on the iterative nonlocal-residual approach proposed by Shaat [27].

In this study, correction factors are formed for local elastic fields of Euler-Bernoulli beams, *i.e.* static bending, natural frequencies, and buckling characteristics. These correction factors are then used to obtain the aforementioned characteristics of nonlocal Euler-Bernoulli beams from their corresponding counterparts of local Euler-Bernoulli beams.

## 2.1 Euler-Bernoulli beams based on the iterative-nonlocal residual approach

The nonlocal residual concept was first presented by Eringen [1,2] and Eringen and Edelen [3] where the sum of the nonlocal interatomic interactions is modeled as a residual field. Then, Shaat [27] was the first to propose the iterative-nonlocal residual approach for the static bending of Euler-Bernoulli beams and Kirchhoff plates. In the context of Shaat's iterative-nonlocal residual approach, the sum of the nonlocal residual at a specific point is computed iteratively. For Euler-Bernoulli beams, in an iteration $k$, the nonlocal stress residual $\tau_{xx}$ is formed as the difference between the local stress field $\sigma_{xx}$ and the nonlocal stress field $t_{xx}$ of the previous iteration, $k-1$, [27]:

$$\tau_{xx}^{(k)} = \sigma_{xx}^{(k-1)} - t_{xx}^{(k-1)} \tag{3}$$

Then, this formed nonlocal stress residual is used to correct the solution of the local field problem to obtain a nonlocal solution. According to this approach, a nonlocal residual-based fictitious force is updated and imposed to the balance equations in each iteration, $k$. For Euler-Bernoulli beams, the equation of motion based on the iterative-nonlocal residual approach can be written in the form [27]:

$$D \frac{\partial^4 w(x,t)^{(k)}}{\partial x^4} + m \frac{\partial^2 w(x,t)^{(k)}}{\partial t^2} = P(x,t) + F(x,t)^{(k)}$$

with

$$D = (\lambda + 2\mu)I \tag{4}$$

where $I$ is the beam's area moment of inertia, and $m$ denotes its mass per unit length. $P(x,t)$ is the applied external force. In equation (4), $F(x,t)$ is introduced as the nonlocal residual-based fictitious force. For Euler-Bernoulli beams, Shaat [27] defined the nonlocal stress residual $\tau_{xx}$ as follows:

$$\tau_{xx}(x,t)^{(k)} = -z(\lambda + 2\mu) \frac{\partial^2 w_c(x,t)^{(k)}}{\partial x^2} \tag{5}$$

and the fictitious force as follows:

$$F(x,t)^{(k)} = D \frac{\partial^4 w_c(x,t)^{(k)}}{\partial x^4} \tag{6}$$

where $w_c(x,t)$ is a deflection-correction field for the nonlocal residual.

For Euler-Bernoulli beams, the nonzero strain component is given by:

$$\varepsilon_{xx}(x,t) = -z \frac{\partial^2 w(x,t)}{\partial x^2} \tag{7}$$

Consequently, by multiplying equation (3) by the nonlocal differential operators, $\left(1 - \epsilon_1 \frac{\partial^2}{\partial x^2}\right)$ and $\left(1 - \epsilon_2 \frac{\partial^2}{\partial x^2}\right)$, and according to equations (1) and (5), the following relation can be obtained for Euler-Bernoulli beams:

$$
\begin{aligned}
&\left(1 - \epsilon_1 \frac{\partial^2}{\partial x^2}\right)\left(1 - \epsilon_2 \frac{\partial^2}{\partial x^2}\right)(\lambda + 2\mu) \frac{\partial^2 w_c(x,t)^{(k)}}{\partial x^2} \\
&= \left\{\left(1 - \epsilon_1 \frac{\partial^2}{\partial x^2}\right)\left(1 - \epsilon_2 \frac{\partial^2}{\partial x^2}\right)(\lambda + 2\mu) - \lambda\left(1 - \epsilon_2 \frac{\partial^2}{\partial x^2}\right) - 2\mu\left(1 - \epsilon_1 \frac{\partial^2}{\partial x^2}\right)\right\} \frac{\partial^2 w(x,t)^{(k-1)}}{\partial x^2}
\end{aligned}
\tag{8}
$$

According to the iterative-nonlocal residual approach, in each iteration, $w_c$ is first obtained from equation (8) and then used to form the fictitious force (equation (6)). This fictitious force is then imposed to the local field problem (equation (4)) where the corrected local fields for the nonlocal residual can be obtained.

For Eringen's nonlocal theory ($\epsilon_1 = \epsilon_2 = \varrho^2$), equation (8) can be reduced to:

$$\left(1 - \xi^2 \frac{\partial^2}{\partial x^2}\right) \frac{\partial^2 w_c(x,t)^{(k)}}{\partial x^2} = \left(-\varrho^2 \frac{\partial^2}{\partial x^2}\right) \frac{\partial^2 w(x,t)^{(k-1)}}{\partial x^2} \tag{9}$$

which is the relation as previously obtained by Shaat [27].

Because they are blended out of the same mold, the beam deflection, $w(x,t)$, and its correction field, $w_c(x,t)$, are decomposed as follows [27]:

$$
\begin{aligned}
w(x,t) &= W\varphi(x)e^{i\omega t} \\
w_c(x,t) &= W_c\varphi(x)e^{i\omega t}
\end{aligned}
\tag{10}
$$

where $W$ denotes the amplitude of the beam deflection, and $W_c$ is the amplitude of the beam deflection-correction field. $\omega$ is the beam's natural frequency.

It should be noted that the shape function, $\varphi(x)$, in equation (10) is the one of the classical-local Euler-Bernoulli beam. This is because of the fact that the equation of motion (equation (4)) is solved in the local field. By substituting equation (10) into equation (8), the following relation between the amplitude of the deflection and the amplitude of its correction field can be obtained:

$$W_c^{(k)} = \psi(x) W^{(k-1)} \tag{11}$$

where

$$\psi(x) = 1 + \frac{\lambda\left(\frac{d^2\varphi(x)}{dx^2} - \epsilon_2\frac{d^4\varphi(x)}{dx^4}\right) + 2\mu\left(\frac{d^2\varphi(x)}{dx^2} - \epsilon_1\frac{d^4\varphi(x)}{dx^4}\right)}{(\lambda + 2\mu)\left[\frac{d^2\varphi(x)}{dx^2} - (\epsilon_1 + \epsilon_2)\frac{d^4\varphi(x)}{dx^4} + \epsilon_1\epsilon_2\frac{d^6\varphi(x)}{dx^6}\right]} \tag{12}$$

Equation (12) presents $\psi(x)$ as a correction parameter which is formed based on the general nonlocal theory. This correction parameter is used to form the fictitious force based on the local elastic field obtained in a previous iteration. This correction parameter can be obtained on the bases of Eringen's nonlocal theory (as obtained by Shaat [27]) by substituting $\epsilon_1 = \epsilon_2 = \varrho^2$ into equation (12), as follows:

$$\psi(x) = \frac{-\varrho^2\frac{d^2\varphi(x)}{dx^2}}{1 - \varrho^2\frac{d^2\varphi(x)}{dx^2}} \tag{13}$$

According to equations (10), (11), and (6), the equation of motion (equation (4)) in an iteration $k$ for Euler-Bernoulli beams can be written in the form:

$$W^{(k)}\left[D\frac{d^4\varphi(x)}{dx^4} - m\omega^2\varphi(x)\right]e^{i\omega t} = P(x,t) + W^{(k-1)}\left[D\frac{d^4\varphi(x)}{dx^4}\psi(x)e^{i\omega t}\right] \tag{14}$$

Equation (14) presents the equation of motion of Euler-Bernoulli beams according to the iterative-nonlocal residual approach.

## 2.2 Static bending of nonlocal Euler-Bernoulli beams

For static bending of Euler-Bernoulli beams, equation (14) can be rewritten in the form:

$$W^{(k)}\left[D\frac{d^4\varphi(x)}{dx^4}\right] = P(x) + W^{(k-1)}\left[D\frac{d^4\varphi(x)}{dx^4}\psi(x)\right] \tag{15}$$

By multiplying equation (15) by $\varphi(x)$ and integrating both sides over the beam length, the beam equilibrium equation for static bending can be obtained in the form:

$$K_L W^{(k)} = \int_0^L P(x)\varphi(x)\,dx + K_C W^{(k-1)} \tag{16}$$

where the stiffnesses, $K_L$ and $K_C$, are obtained in the form:

$$K_L = \int_0^L \left( D \frac{d^4 \varphi(x)}{dx^4} \varphi(x) \right) dx$$

$$K_C = \int_0^L \left( D \frac{d^4 \varphi(x)}{dx^4} \psi(x) \varphi(x) \right) dx \tag{17}$$

where $K_L$ is the local beam stiffness, and $K_C$ is a correction stiffness.

Equation (16) indicates that the local amplitude can be corrected for the nonlocal residual with the incorporation of $K_C W^{(k-1)}$ as a nonlocal residual-based fictitious force. By introducing $R = K_C / K_L$ as a nonlocal residual-based correction factor, equation (16) can be rewritten in the form:

$$W^{(k)} = W^{(1)} + R W^{(k-1)} \text{ with } R = K_C / K_L \tag{18}$$

where

$$W^{(1)} = \frac{\int_0^L \left( p(x) \varphi(x) \right) dx}{K_L} \tag{19}$$

is the amplitude of the local beam deflection.

By introducing $\aleph$ as the number of iterations needed to obtain the nonlocal solution within a relative error $\mathcal{E} = \frac{W^{(\aleph)} - W^{(\aleph-1)}}{W^{(\aleph-1)}}$, the nonlocal amplitude of the beam's deflection becomes:

$$W^{(\aleph)} = W^{(1)} \left( 1 + \sum_{n=1}^{\aleph} R^n \right) \tag{20}$$

where $W^{(\aleph)}$ is the amplitude of the nonlocal beam deflection. Hence, the nonlocal beam deflection can be obtained as follows:

$$w(x)^{(\aleph)} = W^{(1)} \left( 1 + \sum_{q=1}^{\aleph} R^q \right) \varphi(x) \tag{21}$$

According to equation (20), the convergence of equation (18) to the nonlocal solution is guaranteed when the correction factor $R < 1$ [27].

Equations (18) and (21) present an iterative correction for the static bending of Euler-Bernoulli beams for nonlocal residuals. This correction process basically depends on the correction factor, $R$, that depends on the correction parameter $\psi(x)$ (defined in equation (12)). It should be demonstrated that, because the equations of motion are solved in the local fields, the shape function $\varphi(x)$ presented in the previous equations is the classical one of Euler-Bernoulli beams. Thus, the correction factor, $R$, can be easily formed and incorporated in the local prediction/nonlocal correction-type iterative procedure presented in equation (18).

## 2.3  Free vibration of nonlocal Euler-Bernoulli beams

For the free vibration of nonlocal Euler-Bernoulli beams, equation (14) can be written at iteration, $\aleph$, as follows:

$$D\left(\frac{d^4\varphi_n(x)}{dx^4} - \frac{d^4\varphi_n(x)}{dx^4}\psi_n(x)\right) - m\omega_n^2\varphi_n(x) = 0 \text{ , where } W^{(\aleph)} \cong W^{(\aleph-1)} \tag{22}$$

where $\varphi_n(x)$ is the $n$th mode shape function corresponding to the $n$th natural frequency $\omega_n$. For the $n$th mode, $\psi_n(x)$ can be rewritten as follows:

$$\psi_n(x) = 1 + \frac{\lambda\left(\frac{d^2\varphi_n(x)}{dx^2} - \epsilon_2\frac{d^4\varphi_n(x)}{dx^4}\right) + 2\mu\left(\frac{d^2\varphi_n(x)}{dx^2} - \epsilon_1\frac{d^4\varphi_n(x)}{dx^4}\right)}{(\lambda + 2\mu)\left[\frac{d^2\varphi_n(x)}{dx^2} - (\epsilon_1 + \epsilon_2)\frac{d^4\varphi_n(x)}{dx^4} + \epsilon_1\epsilon_2\frac{d^6\varphi_n(x)}{dx^6}\right]} \tag{23}$$

The multiplication of equation (22) by $\varphi_n(x)$ and integrating the result over the beam length give the characteristics equation in the form:

$$(K_L - K_C) - M\omega_n^2 = 0 \tag{24}$$

where

$$M = m\int_0^L \varphi_n(x)\varphi_n(x)\,dx$$

$$K_L = D\int_0^L \frac{d^4\varphi_n(x)}{dx^4}\varphi_n(x)\,dx \tag{25}$$

$$K_C = D\int_0^L \frac{d^4\varphi_n(x)}{dx^4}\psi_n(x)\varphi_n(x)\,dx$$

Consequently, the natural frequencies of the nonlocal beam can be obtained as follows:

$$\omega_n^2 = \left(\frac{K_L - K_C}{M}\right)_n = \Omega_n^2(1 - R_n) \text{ with } R_n = \left(\frac{K_C}{K_L}\right)_n \tag{26}$$

where $\Omega_n$ is the $n$th mode-local natural frequency:

$$\Omega_n^2 = \left(\frac{K_L}{M}\right)_n \tag{27}$$

Equation (26) presents a one-step solution for the nonlocal natural frequencies of Euler-Bernoulli beams. With no iterations, the nonlocal frequencies can be directly obtained for the different modes of vibration from equation (26) utilizing the correction factors, $R_n$.

## 2.4 Buckling characteristics of nonlocal Euler-Bernoulli beams under axial compressive loads

Equation (4), which represents the beam's equation of motion according to the iterative-nonlocal residual approach, can be rewritten for an Euler-Bernoulli beam under an axial compressive load as follows [28,29]:

$$D \frac{\partial^4 w(x,t)^{(k)}}{\partial x^4} + P \frac{\partial^2 w(x,t)^{(k)}}{\partial x^2} + m \frac{\partial^2 w(x,t)^{(k)}}{\partial t^2} = \left[ D \frac{\partial^4 w_c(x,t)^{(k)}}{\partial x^4} \right] \tag{28}$$

where $P$ is the resultant axial load [29]:

$$P = N - \frac{A(\lambda + 2\mu)}{2L} \int_0^L \left( \frac{\partial w(x,t)^{(k)}}{\partial x} \right)^2 dx \tag{29}$$

where $N$ denotes the applied axial compressive load. The second term of equation (29) represents the beams axial stretching-force.

Because of the axial compressive load, the beam buckles when $N > N_{cr}$, *i.e.* $N_{cr}$ is the critical buckling load (beam buckling strength). In this study, the prebuckling ($N < N_{cr}$) and postbuckling ($N > N_{cr}$) characteristics of nonlocal Euler-Bernoulli beams are determined utilizing the iterative-nonlocal residual approach.

### 2.4.1 Prebuckling natural frequencies of nonlocal Euler-Bernoulli beams

An axially compressed Euler-Bernoulli beam maintains its straight configuration when the axial compressive load is less than the beam's buckling strengths, *i.e.* $N < N_{cr}$. At the prebuckling configuration, the natural frequencies of the Euler-Bernoulli beam are significantly influenced by the magnitude of the axial compressive load. In this study, the prebuckling frequencies of Euler-Bernoulli beams are determined. To this end, the beam deflection, $w(x,t)$, and its correction field, $w_c(x,t)$, in the prebuckling configuration are decomposed as presented in equation (10). It should be noted that, in equation (28), $w(x,t)$ represents a disturbance of the beam about its prebuckling configuration.

The substitution of equation (10) into equation (28) leads to:

$$W^{(k)} \left\{ D \left( \frac{d^4 \varphi_n(x)}{dx^4} \right) + P \left( \frac{d^2 \varphi_n(x)}{dx^2} \right) - m\omega^2 \varphi_n(x) \right\} = W_c^{(k)} \left\{ D \left( \frac{d^4 \varphi_n(x)}{dx^4} \right) \right\} \tag{30}$$

with

$$P = N - \left( \frac{A(\lambda + 2\mu)}{2L} \int\limits_0^L \left( W^{(k)} \frac{d\varphi_n(x)}{dx} e^{i\omega t} \right)^2 dx \right)$$

where $W_c$ is related to $W$ via equation (11). Then, by substituting equation (11) into equation (30), dropping the nonlinear term, multiplying the result by $\varphi_n(x)$, and performing an integration over the beam length, the following equation that governs the beam's prebuckling vibration in an iteration $k$ is obtained:

$$W^{(k)}[K_L + K_B - M\omega_n^2] = W^{(k-1)}K_C \qquad (31)$$

where

$$M = m \int_0^L \varphi_n(x)\varphi_n(x)\, dx$$

$$K_L = D \int_0^L \frac{d^4\varphi_n(x)}{dx^4}\varphi_n(x)\, dx$$

$$K_B = N \int_0^L \frac{d^2\varphi_n(x)}{dx^2}\varphi_n(x)\, dx \qquad (32)$$

$$K_C = D \int_0^L \frac{d^4\varphi_n(x)}{dx^4}\psi_n(x)\varphi_n(x)\, dx$$

where $\psi_n(x)$ is defined in equation (23).

Then, the characteristics equation can be obtained by writing equation (31) at iteration $\aleph$; thus,

$$K_L + K_B - K_C - M\omega_n^2 = 0 \text{ , where } W^{(\aleph)} \cong W^{(\aleph-1)} \qquad (33)$$

Hence, the prebuckling natural frequencies of nonlocal Euler-Bernoulli beams can be obtained as follows:

$$\omega_n^2 = \frac{K_L + K_B - K_C}{M} = \Omega_n^2(1 - R_n) \qquad (34)$$

with

$$R_n = \left( \frac{K_C}{K_L + K_B} \right)_n \qquad (35)$$

where $\Omega_n^2 = \left( \frac{K_L + K_B}{M} \right)_n$ denotes the $n$th prebuckling natural frequency of local Euler-Bernoulli beams under axial compressive loads. Utilizing the correction factors $R_n$ defined in equation (35), the natural frequencies of the different modes of the prebuckling vibration of nonlocal Euler-Bernoulli beams under axial compressive loads can be directly obtained using equation (34).

## 2.4.2 Buckling and postbuckling characteristics of nonlocal Euler-Bernoulli beams

The critical buckling loads and the static buckling configurations of nonlocal Euler-Bernoulli beams under axial compressive loads are determined via the proposed iterative-nonlocal residual approach. To this end, the deflection, $w(x, t)$, and its correction field, $w_c(x, t)$, are considered for the $n$th mode of buckling as follows:

$$w(x) = W\varphi_n(x)$$
$$w_c(x) = W_c\varphi_n(x)$$
(36)

where $\varphi_n(x)$ is the normalized beam buckling configuration. $W$ and $W_c$ denote, respectively, the amplitudes of the buckling configuration and its correction field.

The substituting of equation (36) into equation (28) gives the beam's equilibrium equation at its static buckling configuration as follows:

$$W^{(k)} \left\{ D\left(\frac{d^4\varphi_n(x)}{dx^4}\right) + P\left(\frac{d^2\varphi_n(x)}{dx^2}\right) \right\} = W_c^{(k-1)} \left\{ D\left(\frac{d^4\varphi_n(x)}{dx^4}\right) \right\}$$
(37)

with

$$P = N - \frac{A(\lambda + 2\mu)}{2L} \int_0^L \left( W^{(k)} \frac{d\varphi_n(x)}{dx} \right)^2 dx$$
(38)

where $W$ is related to $W_c$ according to equation (11). According to equation (11) and by multiplying equation (37) by $\varphi_n(x)$ and integrating the result over the beam length, the equation governs the buckling configurations of Euler-Bernoulli beams can be written at iteration $\aleph$, *i.e.* $W^{(\aleph)} \cong W^{(\aleph-1)}$, in the form:

$$W^{(\aleph)}[K_L + \lambda_n^2 K_B - K_C] = 0$$
(39)

where

$$K_L = \int_0^L \frac{d^4\varphi_n(x)}{dx^4} \varphi_n(x) \, dx$$

$$K_B = \int_0^L \frac{d^2\varphi_n(x)}{dx^2} \varphi_n(x) \, dx$$
(40)

$$K_C = \int_0^L \frac{d^4\varphi_n(x)}{dx^4} \psi_n(x)\varphi_n(x) \, dx$$

and

$$\lambda_n^2 = \frac{P}{D}$$
(41)

Since $W \neq 0$ when the beam is buckled,

$$[K_L + \lambda_n^2 K_B - K_C] = 0 \tag{42}$$

which gives the eigenvalues $\lambda_n$ that describe the various buckling configurations of Euler-Bernoulli beams. According to equations (41) and (42), the eigenvalues, $\lambda_n$, can be obtained as follows:

$$\lambda_n^2 = \alpha_n^2(1 - R_n) \tag{43}$$

with

$$R_n = \left(\frac{K_C}{K_L}\right)_n \tag{44}$$

where $\alpha_n^2 = \left(-\frac{K_L}{K_B}\right)_n$ are the conventional eigenvalues of local Euler-Bernoulli beams. In equation (43), $R$, is defined as a correction factor to correct the eigenvalues of the local beam for the nonlocal residuals.

In order to form the amplitude of buckling, $W_n$, for the $n$th buckling configuration of the beam, the obtained eigenvalue, $\lambda_n^2$, is substituted into equation (41) which leads to:

$$P = N - W_n^2\left(\frac{A(\lambda + 2\mu)}{2L}\int_0^L \left(\frac{d\varphi_n(x)}{dx}\right)^2 dx\right) = \lambda_n^2 D \tag{45}$$

Consequently:

$$W_n = \pm\sqrt{\left(\frac{2L}{A(\lambda + 2\mu)\int_0^L \left(\frac{d\varphi_n(x)}{dx}\right)^2 dx}\right)[N - \lambda_n^2 D]} \tag{46}$$

where $I$ and $A$ denote, respectively, the beams area moment of inertia and its cross sectional area.

By introducing $\zeta_n = \pm\sqrt{\frac{2L}{A(\lambda + 2\mu)\int_0^L \left(\frac{d\varphi_n(x)}{dx}\right)^2 dx}}(N - \alpha_n^2 D)$, as the amplitude of buckling of local Euler-Bernoulli beams [29], the amplitude of buckling of nonlocal Euler-Bernoulli beams can be defined as follows:

$$W_n = \pm\sqrt{\zeta_n^2 + \left(\frac{2LI\alpha_n^2}{A\int_0^L \left(\frac{d\varphi_n(x)}{dx}\right)^2 dx}\right)R_n} \tag{47}$$

Hence, for the $n$th mode, the corrected beam buckling configuration for the nonlocal residual can be defined as:

$$w(x) = \left( \pm \sqrt{\zeta_n^2 + \left( \frac{2LI\alpha_n^2}{A \int_0^L \left( \frac{d\varphi_n(x)}{dx} \right)^2 dx} \right) R_n} \right) \varphi_n(x) \tag{48}$$

At the initiation of buckling, $W_n = 0$; thus, the critical buckling load $N_{cr}$, *i.e.* the beam buckling strength, can be determined from equation (46) as follows:

$$W_n = 0 \rightarrow N_{cr} = \alpha_n^2 (1 - R_n) D = \mathcal{F}_{cr} (1 - R_n) \tag{49}$$

where $\mathcal{F}_{cr} = \alpha_n^2 D$ denotes the critical buckling load of local Euler-Bernoulli beams.

Equations (48) and (49) represent, respectively, corrections of the beam buckling configuration and critical buckling for the nonlocal residuals via the formation of the correction factor $R$ that is defined in equation (44).

In the previous sections, the correction factors for the static bending, free vibration, prebuckling, buckling, and postbuckling behaviors of Euler-Bernoulli beams are derived. Via these correction factors, the aforementioned quantities and fields of nonlocal Euler-Bernoulli beams can be easily obtained from their counterparts of local Euler-Bernoulli beams. Table 1 shows a summary of the derived correction factors along with the deflections, natural frequencies, prebuckling natural frequencies, buckling configurations, and critical buckling loads of the local and nonlocal Euler-Bernoulli beams.

## 3. Bending, vibration, and buckling of nonlocal simple supported Euler-Bernoulli beams

In this study, we present an easy and effective way to determine the nonlocal fields of Euler-Bernoulli beams from the local fields via some correction factors. These correction factors are formed depending on the normalized mode shape functions of the local Euler-Bernoulli beams. To demonstrate the effectiveness of the proposed iterative-nonlocal residual approach, the static bending, vibration, and buckling characteristics of simple supported nonlocal Euler-Bernoulli beams are determined.

For bending, vibration, and buckling of simple supported-local Euler-Bernoulli beams, the normalized shape functions are determined as follows [27-29]:

$$\varphi_n(x) = \sin \alpha_n x, \textit{ i.e. } \alpha_n = n\pi/L \tag{50}$$

where $n$ is the mode number. For static bending, $n = 1$. $L$ denotes the beam length.

Table 1: Corrections of local elastic fields of Euler-Bernoulli beams for nonlocal residuals: Bending, vibration, and buckling characteristics.

| | |
|---|---|
| $M = m \int_0^L \varphi_n(x) \varphi_n(x)\, dx \;;\; K_L = D \int_0^L \frac{d^4 \varphi_n(x)}{dx^4} \varphi_n(x)\, dx \;;\; K_B = N \int_0^L \frac{d^2 \varphi_n(x)}{dx^2} \varphi_n(x)\, dx \;;$ $K_C = D \int_0^L \frac{d^4 \varphi_n(x)}{dx^4} \psi_n(x) \varphi_n(x)\, dx$ $\psi_n(x) \rightarrow$ equation (23). | |
| Static Bending ($n = 1$) | Local amplitude: $W^{(1)} = \frac{\int_0^L (p(x)\varphi(x))\,dx}{K_L}$ Local deflection: $w(x)^{(1)} = W^{(1)} \varphi(x)$ Correction factor: $R = K_C / K_L$ Nonlocal amplitude: $W^{(\aleph)} = W^{(1)} \left(1 + \sum_{q=1}^{\aleph} R^q \right)$ Nonlocal deflection: $w(x)^{(\aleph)} = W^{(\aleph)} \varphi(x)$ |
| Free Vibration | Local natural frequencies: $\Omega_n = \sqrt{\left(\frac{K_L}{M}\right)_n}$ Correction factors: $R_n = \left(\frac{K_C}{K_L}\right)_n$ Nonlocal natural frequencies: $\omega_n = \sqrt{\Omega_n^2 (1 - R_n)}$ |
| Prebuckling Frequencies | Local natural frequencies: $\Omega_n = \sqrt{\left(\frac{K_L + K_B}{M}\right)_n}$ Correction factors: $R_n = \left(\frac{K_C}{K_L + K_B}\right)_n$ Nonlocal natural frequencies: $\omega_n = \sqrt{\Omega_n^2 (1 - R_n)}$ |
| Buckling Configurations | Local eigenvalues: $\alpha_n^2 = \left(-\frac{K_L}{K_B}\right)_n$ Local amplitudes: $\zeta_n = \pm \sqrt{\frac{2LI}{AD \int_0^L \left(\frac{d\varphi_n(x)}{dx}\right)^2 dx} \left(N - \alpha_n^2 D\right)}$ Local buckling configurations: $w_n(x) = \zeta_n \varphi_n(x)$ Correction factors: $R_n = \left(\frac{K_C}{K_L}\right)_n$ Nonlocal eigenvalues: $\lambda_n^2 = \alpha_n^2 (1 - R_n)$ Nonlocal amplitudes: $W_n = \pm \sqrt{\zeta_n^2 + \left(\frac{2LI\alpha_n^2}{A \int_0^L \left(\frac{d\varphi_n(x)}{dx}\right)^2 dx}\right) R_n}$ Nonlocal buckling configurations: $w_n(x) = W_n \varphi_n(x)$ |
| Critical Buckling | Local eigenvalues: $\alpha_n^2 = \left(-\frac{K_L}{K_B}\right)_n$ Local critical buckling loads: $\mathcal{F}_{cr} = \alpha_n^2 D$ Correction factors: $R_n = \left(\frac{K_C}{K_L}\right)_n$ Nonlocal critical buckling loads: $N_{cr} = \mathcal{F}_{cr}(1 - R_n)$ |

By substituting equation (50) into equation (23), the correction parameter, $\psi_n(x)$, is obtained as follows:

$$\psi_n = 1 - \frac{\lambda(\epsilon_2 \alpha_n^2 + 1) + 2\mu(\epsilon_1 \alpha_n^2 + 1)}{(\lambda + 2\mu)(1 + (\epsilon_1 + \epsilon_2)\alpha_n^2 + \epsilon_1 \epsilon_2 \alpha_n^4)} \tag{51}$$

As presented in equation (51), for simple supported Euler-Bernoulli beams, $\psi$, is a correction constant. According to the general nonlocal theory [19,20], this correction constant depends on two nonlocal parameters, $\epsilon_1$ and $\epsilon_2$. When $\epsilon_1 = \epsilon_1 = \varrho^2$, the correction constant, $\psi$, can be formed according to Eringen's nonlocal theory as follows [27]:

$$\psi_n = 1 - \frac{1}{1 + \varrho^2 \alpha_n^2} \tag{52}$$

The static bending, free vibration, and buckling of nonlocal Euler-Bernoulli beams can be easily determined according to Table 1. Thus, the quantities $M$, $K_L$, $K_B$, $K_C$ are first obtained. Then, the correction factors are formed to correct the local fields for the nonlocal residuals. Table 2 shows the corrections of the local simple supported Euler-Bernoulli beams for the nonlocal residuals.

### 3.1 Static bending

For static bending of simple supported Euler-Bernoulli beams, the correction factor is obtained by $R = \psi_1$. Utilizing this correction factor, the deflection of local Euler-Bernoulli beams can be corrected for the nonlocal residual in a few iterations. For more details on the iterative procedure, readers are referred to Shaat [27]. The results of the proposed approach for the static bending of simple supported Euler-Bernoulli beams subjected to uniform loads are shown in Figures 1 and 2 and Table 3. Figure 1 shows the nondimensional central deflection, $\bar{w}(L/2) = w(L/2)(D/qL^4)$, as a function of the nondimensional nonlocal parameter, $\sqrt{\epsilon_2}/L$, for different $\epsilon_1/\epsilon_2$ ratios when the elastic moduli are considered such that $\lambda = \mu$. The plotted curves in Figure 2, on the other hand, reflect the effects of the elastic moduli ratio, $\frac{\lambda}{\mu} = \frac{2\nu}{1-2\nu}$ where $\nu$ is the Poisson's ratio, on the nondimensional central deflection. The depicted results in Figures 1 and 2 are numerically represented in Table 3.

Table 2: Corrections of the bending, vibration, and buckling characteristics of local simple supported Euler-Bernoulli beams for nonlocal residuals.

| | |
|---|---|
| $\alpha_n = \frac{n\pi}{L}$; $M = \frac{mL}{2}$; $K_L = \frac{DL}{2}(n\pi/L)^4$; $K_B = -\frac{NL}{2}(n\pi/L)^2$; $\psi_n = 1 - \frac{\lambda(\epsilon_2(n\pi/L)^2+1)+2\mu(\epsilon_1(n\pi/L)^2+1)}{(\lambda+2\mu)(1+(\epsilon_1+\epsilon_2)(n\pi/L)^2+\epsilon_1\epsilon_2(n\pi/L)^4)}$; $K_C = \frac{DL}{2}(n\pi/L)^4\psi_n$ | |
| Static Bending (uniform load), *i.e.* $q_0$ load intensity. | Local amplitude: $W^{(1)} = \frac{4q_0}{\pi D}(L/\pi)^4$ |
| | Local deflection: $w(x)^{(1)} = \frac{4q_0}{\pi D}(L/\pi)^4 \sin\frac{\pi}{L}x$ |
| | Correction factor: $R = \psi_1$ |
| | Nonlocal amplitude: $W^{(\aleph)} = \frac{4q_0}{\pi D}(L/\pi)^4\left(1+\sum_{q=1}^{\aleph}R^q\right)$ |
| | Nonlocal deflection: $w(x)^{(\aleph)} = \frac{4q_0}{\pi D}(L/\pi)^4\left(1+\sum_{q=1}^{\aleph}R^q\right)\sin\frac{\pi}{L}x$ |
| Free Vibration | Local natural frequencies: $\Omega_n = \sqrt{\frac{D}{m}(n\pi/L)^4}$ |
| | Correction factors: $R_n = \psi_n$ |
| | Nonlocal natural frequencies: $\omega_n = \sqrt{\frac{D}{m}(n\pi/L)^4(1-R_n)}$ |
| Prebuckling Frequencies | Local natural frequencies: $\Omega_n = \sqrt{(n\pi/L)^2\left[\frac{D(n\pi/L)^2-N}{m}\right]}$ |
| | Correction factors: $R_n = \frac{\left(\frac{n\pi}{L}\right)^2\psi_n}{\left(\frac{n\pi}{L}\right)^2-\frac{N}{D}}$ |
| | Nonlocal natural frequencies: $\omega_n = \sqrt{(n\pi/L)^2(1-R_n)\left[\frac{D(n\pi/L)^2-N}{m}\right]}$ |
| Buckling Configurations | Local amplitudes: $\zeta_n = \pm\sqrt{\frac{4I}{A}\left(\frac{N}{(n\pi/L)^2 D}-1\right)}$ |
| | Local buckling configurations: $w_n(x) = \pm\sqrt{\frac{4I}{A}\left(\frac{N}{(n\pi/L)^2 D}-1\right)}\sin\frac{n\pi}{L}x$ |
| | Correction factors: $R_n = \psi_n$ |
| | Nonlocal eigenvalues: $\lambda_n^2 = (n\pi/L)^2(1-R_n)$ |
| | Nonlocal amplitudes: $W_n = \pm\sqrt{\frac{4I}{A}\left(\frac{N}{(n\pi/L)^2 D}+R_n-1\right)}$ |
| | Nonlocal buckling configurations: $w_n(x) = \pm\sqrt{\frac{4I}{A}\left(\frac{N}{(n\pi/L)^2 D}+R_n-1\right)}\sin\frac{n\pi}{L}x$ |
| Critical Buckling | Local critical buckling loads: $\mathcal{F}_{cr} = (n\pi/L)^2 D$ |
| | Correction factors: $R_n = \psi_n$ |
| | Nonlocal critical buckling loads: $N_{cr} = (n\pi/L)^2 D(1-R_n)$ |

To demonstrate the effectiveness of the proposed corrections of the static bending of local beams for the nonlocal residuals, the results of the proposed approach are plotted parallel to the results of the conventional nonlocal solutions of Eringen's nonlocal model ($\epsilon_1 = \epsilon_2$) [18] in Figure 1. It is clear that the proposed approach can effectively reflect the exact results as the conventional solutions of the nonlocal elastic field problem.

Figure 1 and Table 3 reflect the influences of the nonlocal residuals on the static bending of Euler-Bernoulli beams. Because of these nonlocal residuals, the beam is subjected to a softening mechanism where the increase in the nondimensional nonlocal parameter, $\sqrt{\epsilon_2}/L$, is accompanied with an increase in the beam's nondimensional deflection. Moreover, the severity of the beam's nonlocal softening increases with the increase in $\epsilon_1/\epsilon_2$ ratio, as observed in Figure 1 and Table 3. Eringen's nonlocal theory, *i.e.* $\epsilon_1/\epsilon_2 = 1$, can only model moderate nonlocal fields. However, the general nonlocal theory can effectively model weak (*i.e.* $\epsilon_1/\epsilon_2 < 1$) and strong nonlocal fields (*i.e.* $\epsilon_1/\epsilon_2 > 1$) as well as moderate nonlocal fields (*i.e.* $\epsilon_1/\epsilon_2 = 1$). Shaat [19, 20] has demonstrated that, to accurately model the nonlocal residuals of materials, two distinct nonlocal parameters may be needed, and for such a case, the general nonlocal theory should be employed. Thus, for these materials, Eringen's nonlocal theory may lead to under/over estimations of their nonlocal residuals. The general nonlocal theory, however, presents a remedy for Eringen's nonlocal theory where it gives more freedom to effectively model the nonlocal fields of materials.

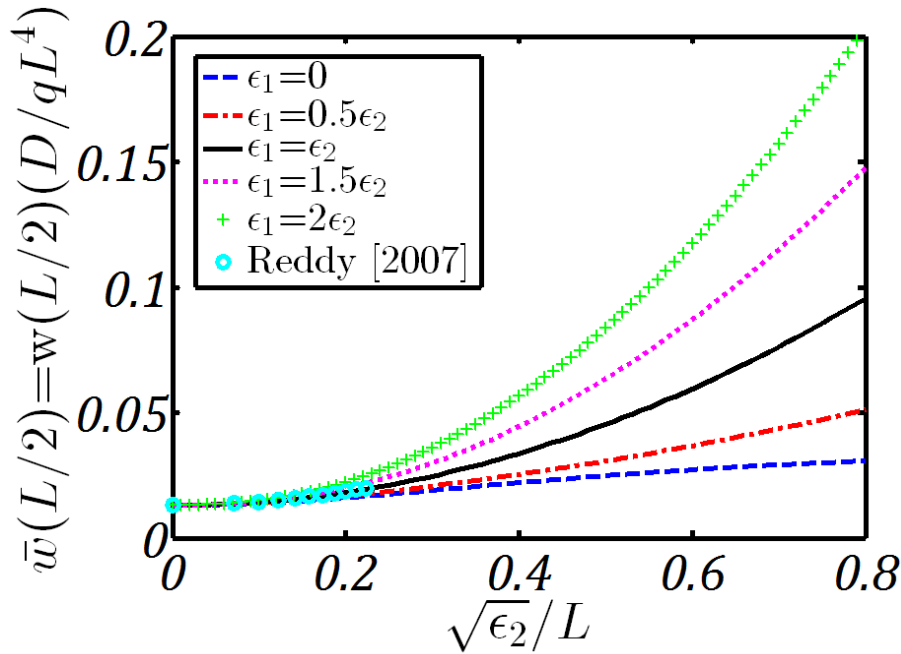

Figure 1: Nondimensional central deflection, $\bar{w}(L/2) = w(L/2)(D/qL^4)$, of simple supported nonlocal Euler-Bernoulli beams as a function of the nondimensional nonlocal parameter, $\sqrt{\epsilon_2}/L$, ($\lambda = \mu$, *i.e.* $\nu = 0.25$).

Table 3: Nondimensional central deflections $\bar{w}(L/2) = w(L/2)(D/qL^4)$ of simple supported Euler-Bernoulli beams (*uniform load*).

| $\sqrt{\epsilon_2}/L$ | $\epsilon_1 = 0$ | $\epsilon_1 = 0.5\epsilon_2$ | $\epsilon_1 = \epsilon_2$ | $\epsilon_1 = 1.5\epsilon_2$ | $\epsilon_1 = 2\epsilon_2$ |
|---|---|---|---|---|---|
| $\lambda = 0.5\mu$, *i.e.* $\nu = 0.1667$ | | | | | |
| 0 | 0.0131 | 0.0131 | 0.0131 | 0.0131 | 0.0131 |
| 0.1 | 0.0141 | 0.0142 | 0.0144 | 0.0146 | 0.0148 |
| 0.2 | 0.0169 | 0.0172 | 0.0182 | 0.0197 | 0.0215 |
| 0.3 | 0.0210 | 0.0216 | 0.0247 | 0.0291 | 0.0343 |
| 0.4 | 0.0256 | 0.0268 | 0.0337 | 0.0430 | 0.0534 |
| 0.5 | 0.0303 | 0.0328 | 0.0453 | 0.0612 | 0.0786 |
| 0.6 | 0.0348 | 0.0396 | 0.0595 | 0.0837 | 0.1098 |
| 0.7 | 0.0388 | 0.0472 | 0.0762 | 0.1105 | 0.1467 |
| 0.8 | 0.0422 | 0.0558 | 0.0956 | 0.1414 | 0.1895 |
| $\lambda = \mu$, *i.e.* $\nu = 0.25$ | | | | | |
| 0 | 0.0131 | 0.0131 | 0.0131 | 0.0131 | 0.0131 |
| 0.1 | 0.0139 | 0.0141 | 0.0144 | 0.0147 | 0.0150 |
| 0.2 | 0.0161 | 0.0169 | 0.0182 | 0.0200 | 0.0221 |
| 0.3 | 0.0190 | 0.0208 | 0.0247 | 0.0299 | 0.0359 |
| 0.4 | 0.0221 | 0.0254 | 0.0337 | 0.0444 | 0.0565 |
| 0.5 | 0.0249 | 0.0308 | 0.0453 | 0.0636 | 0.0837 |
| 0.6 | 0.0272 | 0.0368 | 0.0595 | 0.0872 | 0.1173 |
| 0.7 | 0.0292 | 0.0436 | 0.0762 | 0.1152 | 0.1572 |
| 0.8 | 0.0308 | 0.0513 | 0.0956 | 0.1477 | 0.2033 |
| $\lambda = 1.5\mu$, *i.e.* $\nu = 0.3$ | | | | | |
| 0 | 0.0131 | 0.0131 | 0.0131 | 0.0131 | 0.0131 |
| 0.1 | 0.0138 | 0.0140 | 0.0144 | 0.0147 | 0.0151 |
| 0.2 | 0.0156 | 0.0166 | 0.0182 | 0.0203 | 0.0226 |
| 0.3 | 0.0179 | 0.0202 | 0.0247 | 0.0305 | 0.0372 |
| 0.4 | 0.0201 | 0.0245 | 0.0337 | 0.0455 | 0.0590 |
| 0.5 | 0.0220 | 0.0294 | 0.0453 | 0.0653 | 0.0877 |
| 0.6 | 0.0236 | 0.0350 | 0.0595 | 0.0898 | 0.1233 |
| 0.7 | 0.0248 | 0.0414 | 0.0762 | 0.1189 | 0.1656 |
| 0.8 | 0.0258 | 0.0485 | 0.0956 | 0.1526 | 0.1656 |

To investigate the coupled effects of the nonlocal parameter and the material's Poisson's ratio, $\nu = \lambda/2(\lambda + \mu)$, on the static bending of Euler-Bernoulli beams, the nondimensional central deflections are presented as functions of the elastic moduli ratio, $\lambda/\mu$, for different $\epsilon_1/\epsilon_2$ ratios. The deflections of the local beam ($\epsilon_1 = \epsilon_2 = 0$) are presented by a doted horizontal line, as shown in Figure 2. It is clear that, when the nonlocal residuals are considered, higher nondimensional central deflections are obtained. The dashed horizontal line in Figure 2 represents the results of Eringen's nonlocal model ($\epsilon_1 = \epsilon_2$). It follows from this horizontal line that Eringen's nonlocal model works independently from the material's Poisson's ratio. On the other hand, the effects of the material's Poisson's ratio on the sum of the nonlocal residuals can be depicted via the general nonlocal theory. Thus, the general nonlocal theory can reflect the different severities of the nonlocal residuals on the beam's static deflection, as observed in Figure 2 and Table 3. When $\epsilon_1/\epsilon_2 < 1$, the increase in the elastic moduli ratio, $\lambda/\mu$, is accompanied with a decrease in the nondimensional central deflection. On the other hand, opposite trends are obtained when $\epsilon_1/\epsilon_2 > 1$.

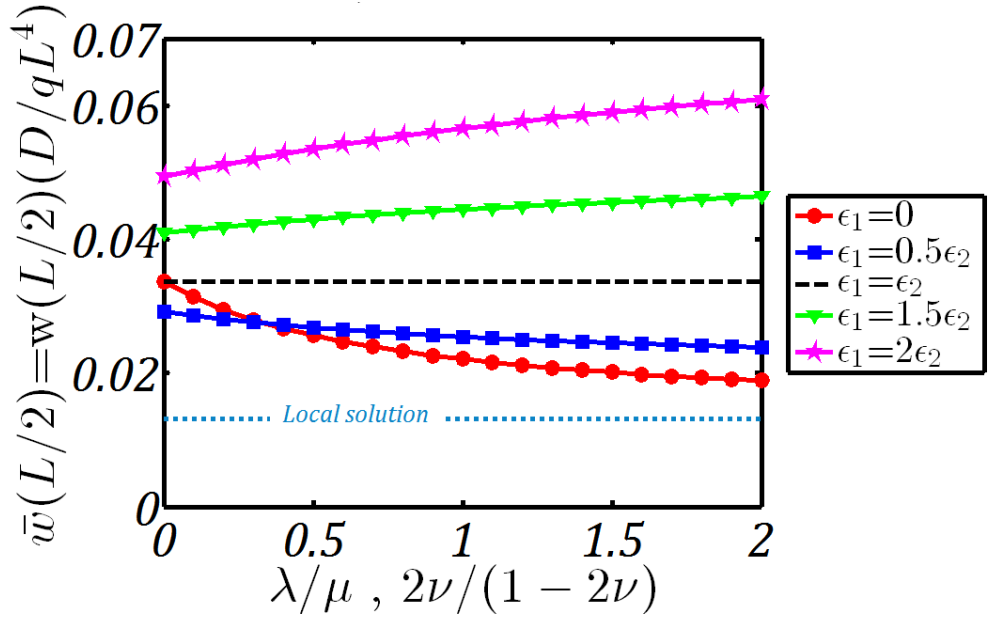

Figure 2: Nondimensional central deflection, $\bar{w}(L/2) = w(L/2)(D/qL^4)$, of simple supported nonlocal Euler-Bernoulli beams as a function of elastic moduli ratio, $\lambda/\mu$, ($\sqrt{\epsilon_2}/L = 0.4$).

## 3.2 Free vibration

To determine the natural frequencies of simple supported nonlocal Euler-Bernoulli beams, the correction factors for the different modes are obtained as $R_n = \psi_n$. The correction factors of the first four modes of vibration are plotted as functions of the nondimensional nonlocal parameter, $\sqrt{\epsilon_2}/L$, and the elastic moduli ratio, $\lambda/\mu$, in Figures 3 and 4, respectively.

Figure 3 reflects the increase in the correction factors with the increase in the nondimensional nonlocal parameter, $\sqrt{\epsilon_2}/L$, and/or the nonlocal parameters-ratio, $\epsilon_1/\epsilon_2$. Because the shifts in the natural frequencies from their counterparts of the local beam increase with the increase in the nondimensional nonlocal parameters, higher correction factors are needed. Moreover, it is clear that zero correction factors are obtained when $\epsilon_1 = \epsilon_2 = 0$ where no corrections should be involved. Moreover, at a certain nonlocal parameter, higher correction factors are needed to obtain the natural frequencies of the higher modes. Inspecting Figure 3, it is shown that the correction factors of the different modes deviate within the range $0 \leq R_n \leq 1$. As previously mentioned, to guarantee the convergence of the iterative approach to the nonlocal solution, the correction factors should be within the aforementioned range.

The impacts of the elastic moduli ratio $\lambda/\mu$ (or Poisson's ratio) on the correction factors of the free vibration of simple supported Euler-Bernoulli beams are depicted in Figure 4. It is clear that, when $\epsilon_1/\epsilon_2 = 1$ (Eringen's model), the correction factors are obtained independent from the elastic moduli ratio $\lambda/\mu$ (the dashed horizontal lines). On the other hand, the correction factors may increase or decrease with the increase in the elastic moduli ratio when $\epsilon_1/\epsilon_2 \neq 1$. Thus, when $\epsilon_1/\epsilon_2 > 1$, the correction factors increase with the increase in the elastic moduli ratio $\lambda/\mu$. An opposite trend, however, is observed when $\epsilon_1/\epsilon_2 < 1$.

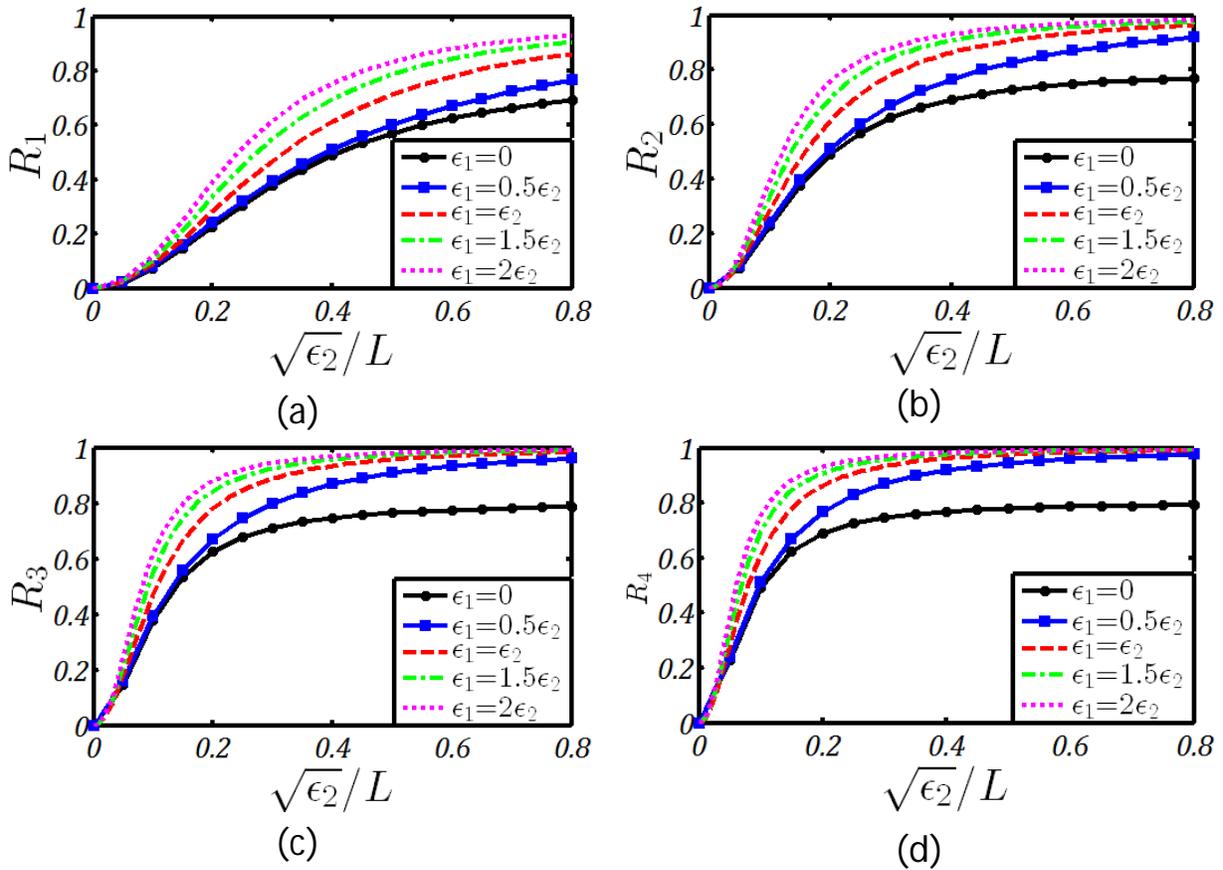

Figure 3: The correction factors of the first four modes of free vibration of simple supported Euler-Bernoulli beams as functions of the nondimensional nonlocal parameter $\sqrt{\epsilon_2}/L$ ($\lambda = 0.5\mu$, *i.e.* $\nu = 0.1667$).

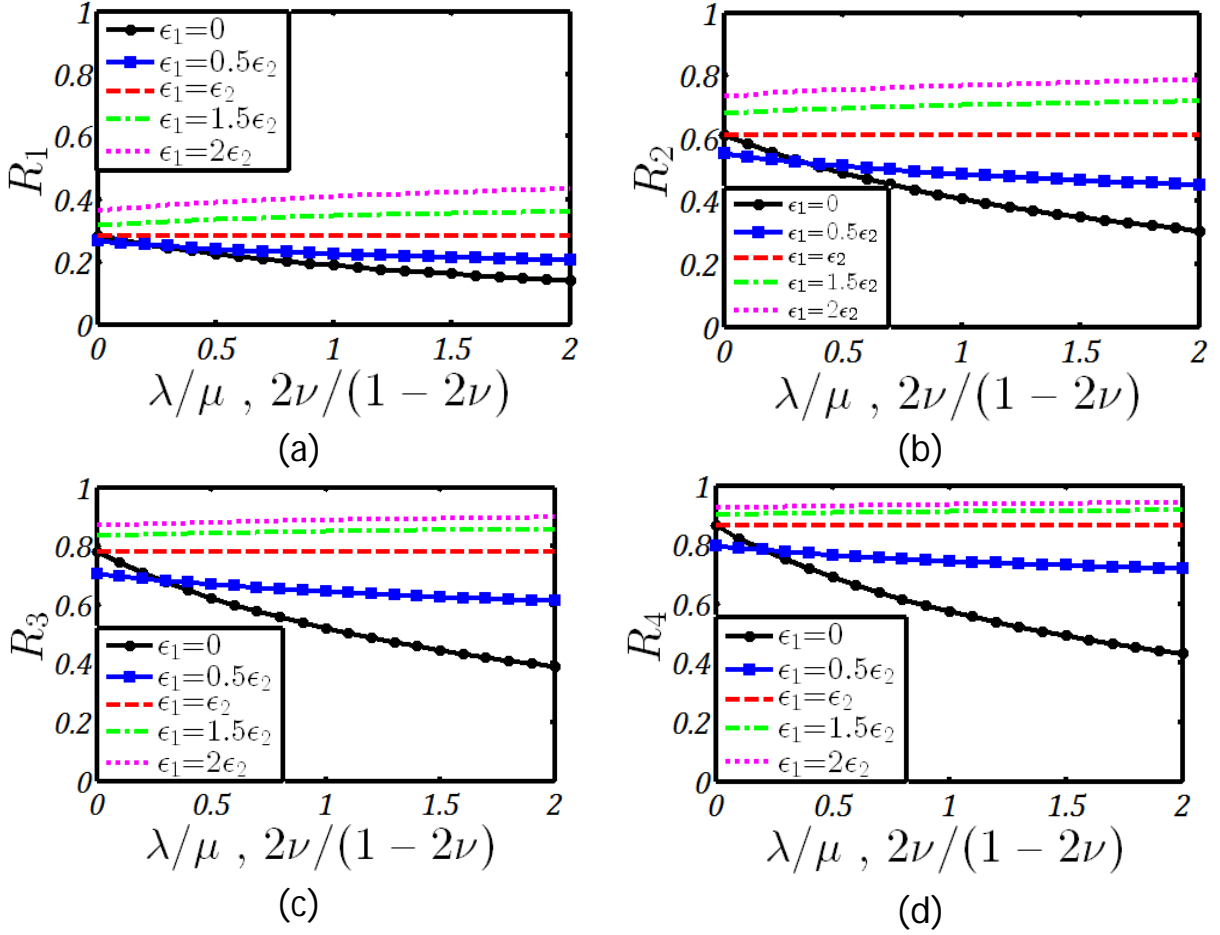

Figure 4: The correction factors of the first four modes of free vibration of simple supported Euler-Bernoulli beams as functions of the elastic moduli ratio $\lambda/\mu$ ($\sqrt{\epsilon_2}/L = 0.2$).

Utilizing the obtained correction factors, the nondimensional natural frequencies, $\bar{\omega}_n = \omega_n\sqrt{mL^4/D}$, of the first four modes of the free-vibration of simple supported nonlocal Euler-Bernoulli beams are depicted as functions of the nondimensional nonlocal parameter, $\sqrt{\epsilon_2}/L$, and the elastic moduli ratio, $\lambda/\mu$, in Figures 5 and 6, respectively. To demonstrate the effectiveness of the proposed approach to determine the natural frequencies of nonlocal beams, the nondimensional natural frequencies as obtained via the proposed approach are compared to the ones determined by Lu et al. [14] when $\epsilon_1 = \epsilon_2$, as presented in Figure 5. The results of the proposed approach perfectly fit the results determined by Lu et al. [14]. However, ease of determining the nonlocal frequencies is the main merit of the proposed approach over the existing solutions of the nonlocal field problem.

It follows from Figure 5 that the increase in the nondimensional nonlocal parameter is accompanied with reductions in the nondimensional natural frequencies. Moreover, at a certain value of the nondimensional nonlocal parameter, the nondimensional natural frequencies decrease with the increase in the nonlocal parameters-ratio, $\epsilon_1/\epsilon_2$. This demonstrates that, for the considered nonlocal parameters, the nonlocal theory affects the beam with a softening mechanism. Moreover, Figure 5 shows that the general nonlocal theory can give the same results as Eringen's nonlocal theory (which only assumes $\epsilon_1 = \epsilon_2$). The general nonlocal theory, however, outperforms Eringen's nonlocal theory for considering the significant effects of the nonlocal parameters-ratio. Thus, big discrepancies between the two theories exist when $\epsilon_1/\epsilon_2 \neq 1$. Thus, Eringen's model can only be used when $\epsilon_1/\epsilon_2 = 1$; otherwise, the general nonlocal theory should be employed.

In Figure 6, the impacts of the elastic moduli ratio (or Poisson's ratio) on the nondimensional natural frequencies of simple supported Euler-Bernoulli beams are depicted. Inspecting the plotted curves in Figure 6, it is clear that Eringen's model (*i.e.* $\epsilon_1 = \epsilon_2$) gives the same nondimensional natural frequencies for all the material moduli values. However, the dependency of the nondimensional natural frequencies on the beam's elastic moduli and Poisson's ratio can only be depicted via the general nonlocal theory. Utilizing the general nonlocal theory, when $\epsilon_1/\epsilon_2 < 1$, the reductions in the nondimensional natural frequencies with the increase in the elastic moduli ratio are observed in Figure 6; on the other hand, opposite trends are obtained when $\epsilon_1/\epsilon_2 < 1$.

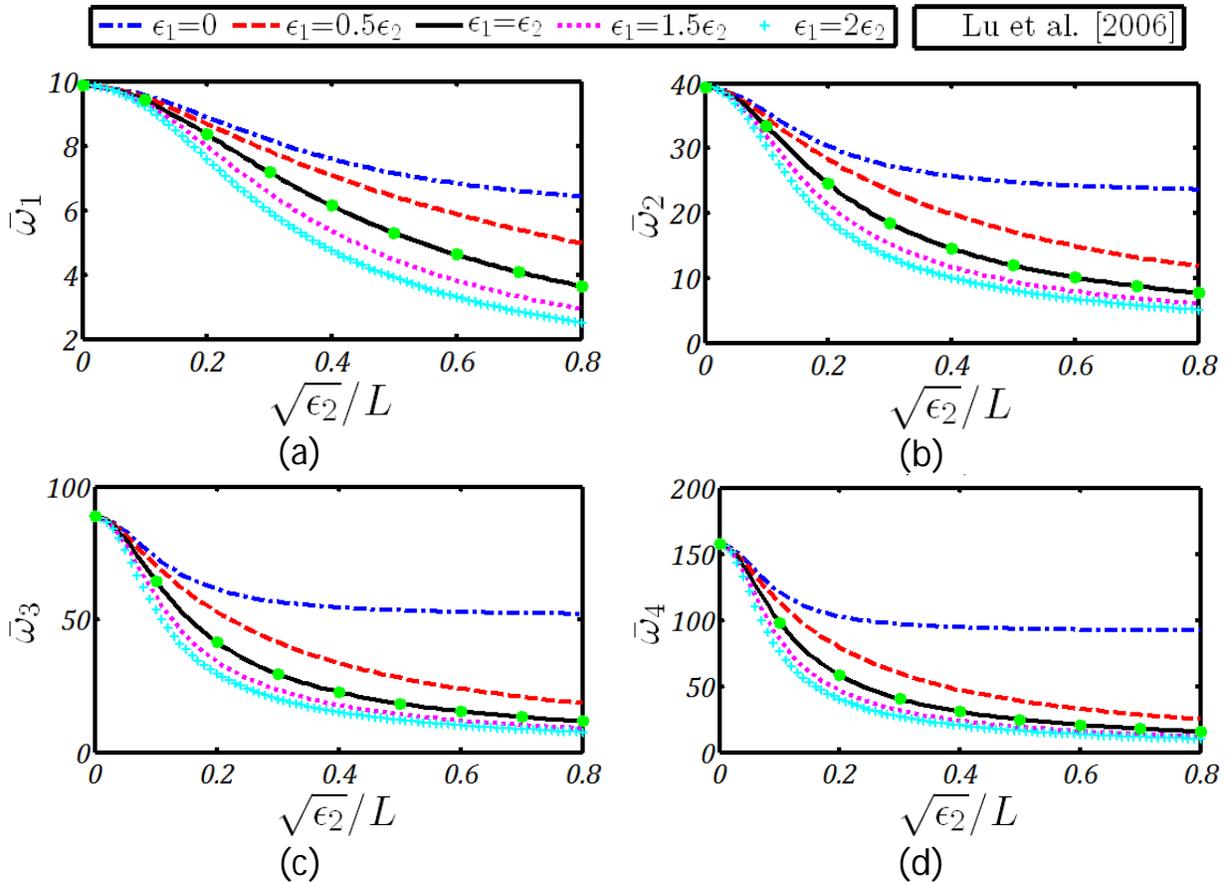

Figure 5: Nondimensional natural frequencies of the first four modes of vibration of simple supported nonlocal Euler-Bernoulli beams as functions of the nondimensional nonlocal parameter, $\sqrt{\epsilon_2}/L$, ($\lambda/\mu = 1$, *i.e.* $\nu = 0.25$).

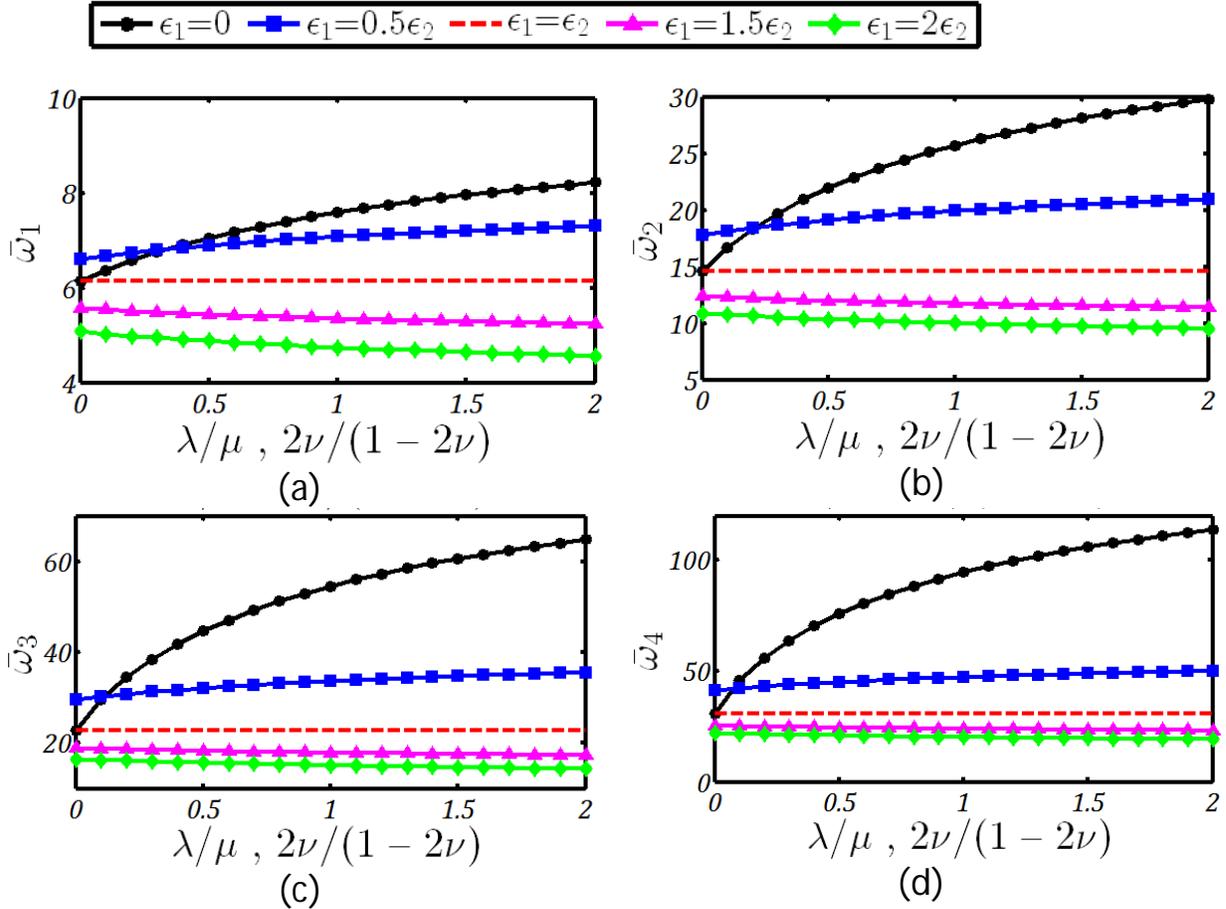

Figure 6: Nondimensional natural frequencies of the first four modes of vibration of simple supported nonlocal Euler-Bernoulli beams as functions of the elastic moduli ratio, $\lambda/\mu$, ($\sqrt{\epsilon_2}/L = 0.4$).

### 3.3 Critical buckling loads

To determine the critical buckling loads of simple supported nonlocal Euler-Bernoulli beams, correction factors are defined for the different modes in Table 2 by $R_n = \psi_n$. These correction factors are harnessed to directly determine the critical buckling loads of nonlocal beams from their counterparts of the local beams. It should be mentioned that these correction factors are exactly the same as those of the free vibration of beams. Thus, the same correction factors for the different modes of buckling as the ones presented in Figure 4 can be obtained. The critical buckling loads for simple supported Euler-Bernoulli beams are reported in in Table 4 and Figure 7. To demonstrate that the proposed approach can give the same results as the ones of the existing solutions of Eringen's model, a comparison with the results obtained by Reddy [18] is performed, as presented in Figure 7. The accuracy and effectiveness of the proposed approach is obvious where the exact same results as those presented in Reddy [18] are obtained. It follows from Table 4 and Figure 7 that the increase in the nondimensional nonlocal parameter, $\sqrt{\epsilon_2}/L$, and/or the nonlocal parameters-ratio, $\epsilon_1/\epsilon_2$, is accompanied with a decrease in the nondimensional critical buckling load.

Table 4: Nondimensional critical buckling loads, $\bar{N}_{cr} = N_{cr}L^2/\pi^2 D$, of simple supported nonlocal Euler-Bernoulli beams.

| $\sqrt{\epsilon_2}/L$ | $\epsilon_1 = 0$ | $\epsilon_1 = 0.5\epsilon_2$ | $\epsilon_1 = \epsilon_2$ | $\epsilon_1 = 1.5\epsilon_2$ | $\epsilon_1 = 2\epsilon_2$ |
|---|---|---|---|---|---|
| $\lambda = 0.5\mu$, *i.e.* $\nu = 0.1667$ | | | | | |
| 0 | 1 | 1 | 1 | 1 | 1 |
| 0.1 | 0.9281 | 0.9207 | 0.9102 | 0.8971 | 0.8821 |
| 0.2 | 0.7736 | 0.7583 | 0.7170 | 0.6642 | 0.6092 |
| 0.3 | 0.6237 | 0.6060 | 0.5296 | 0.4491 | 0.3810 |
| 0.4 | 0.5102 | 0.4878 | 0.3877 | 0.3041 | 0.2445 |
| 0.5 | 0.4307 | 0.3986 | 0.2884 | 0.2135 | 0.1661 |
| 0.6 | 0.3757 | 0.3302 | 0.2196 | 0.1560 | 0.1190 |
| 0.7 | 0.3371 | 0.2767 | 0.1713 | 0.1182 | 0.0890 |
| 0.8 | 0.3093 | 0.2342 | 0.1367 | 0.0924 | 0.0689 |
| $\lambda = \mu$, *i.e.* $\nu = 0.25$ | | | | | |
| 0 | 1 | 1 | 1 | 1 | 1 |
| 0.1 | 0.9401 | 0.9264 | 0.9102 | 0.8920 | 0.8722 |
| 0.2 | 0.8113 | 0.7744 | 0.7170 | 0.6530 | 0.5905 |
| 0.3 | 0.6864 | 0.6294 | 0.5296 | 0.4372 | 0.3637 |
| 0.4 | 0.5918 | 0.5142 | 0.3877 | 0.2941 | 0.2312 |
| 0.5 | 0.5256 | 0.4250 | 0.2884 | 0.2056 | 0.1561 |
| 0.6 | 0.4798 | 0.3551 | 0.2196 | 0.1499 | 0.1114 |
| 0.7 | 0.4476 | 0.2996 | 0.1713 | 0.1133 | 0.0831 |
| 0.8 | 0.4245 | 0.2548 | 0.1367 | 0.0884 | 0.0642 |
| $\lambda = 1.5\mu$, *i.e.* $\nu = 0.3$ | | | | | |
| 0 | 1 | 1 | 1 | 1 | 1 |
| 0.1 | 0.9487 | 0.9305 | 0.9102 | 0.8883 | 0.8652 |
| 0.2 | 0.8383 | 0.7859 | 0.7170 | 0.6449 | 0.5771 |
| 0.3 | 0.7312 | 0.6461 | 0.5296 | 0.4288 | 0.3513 |
| 0.4 | 0.6501 | 0.5330 | 0.3877 | 0.2870 | 0.2216 |
| 0.5 | 0.5934 | 0.4439 | 0.2884 | 0.2000 | 0.1489 |
| 0.6 | 0.5541 | 0.3730 | 0.2196 | 0.1454 | 0.1059 |
| 0.7 | 0.5265 | 0.3159 | 0.1713 | 0.1098 | 0.0788 |
| 0.8 | 0.5067 | 0.2695 | 0.1367 | 0.0856 | 0.0609 |

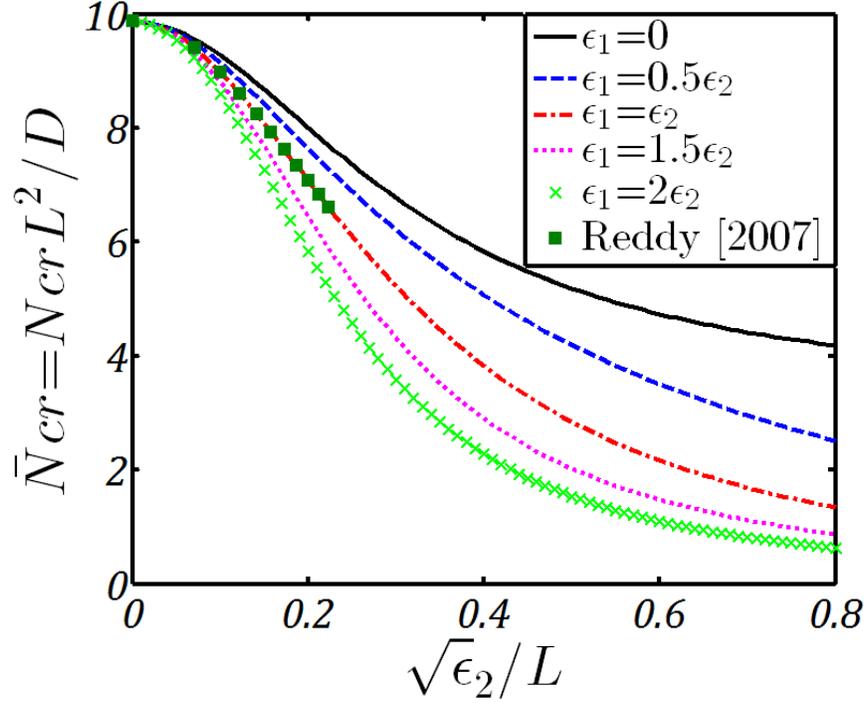

Figure 7: The nondimensional critical buckling load, $\bar{N}_{cr} = N_{cr}L^2/D$, as a function of the nondimensional nonlocal parameter, $\sqrt{\epsilon_2}/L$, ($\lambda = \mu$, *i.e.* $\nu = 0.25$).

### 3.4 Prebuckling natural frequencies

The prebuckling natural frequencies of simple supported nonlocal Euler-Bernoulli beams under axial compressive loads are investigated. The correction factors for the different modes of the prebuckling vibration are determined by $R_n = \left(\left(\frac{n\pi}{L}\right)^2 \psi_n\right) \Big/ \left(\left(\frac{n\pi}{L}\right)^2 - \frac{N}{D}\right)$, as presented in Table 2. These correction factors not only depend on the nonlocal parameters but also depend on the applied axial compressive load, $N$. To demonstrate this fact, the correction factors of the prebuckling-fundamental natural frequency as functions of the nondimensional nonlocal parameter, $\sqrt{\epsilon_2}/L$, and the nondimensional axial compressive load, $\bar{N} = NL^2/\pi^2 D$, are plotted in Figure 8. The plotted curves in Figure 8 reflect the increase in the correction factors with the increase in the nondimensional axial load and/or the nondimensional nonlocal parameter. Moreover, at certain nonlocal parameter and axial load value, the increase in the nonlocal parameters-ratio, $\epsilon_1/\epsilon_2$, is accompanied with an increase in the correction factor. It should be mentioned that the correction factor, $R_1$, reaches its maximum value, *i.e.* $R_1 = 1$, when the beam is subjected to its critical buckling load, *i.e.* $N = N_{cr}$. Thus, for $N < N_{cr}$, the correction factor is confined within the range $0 \leq R_1 < 1$ depending on the elastic moduli and the nonlocal parameters.

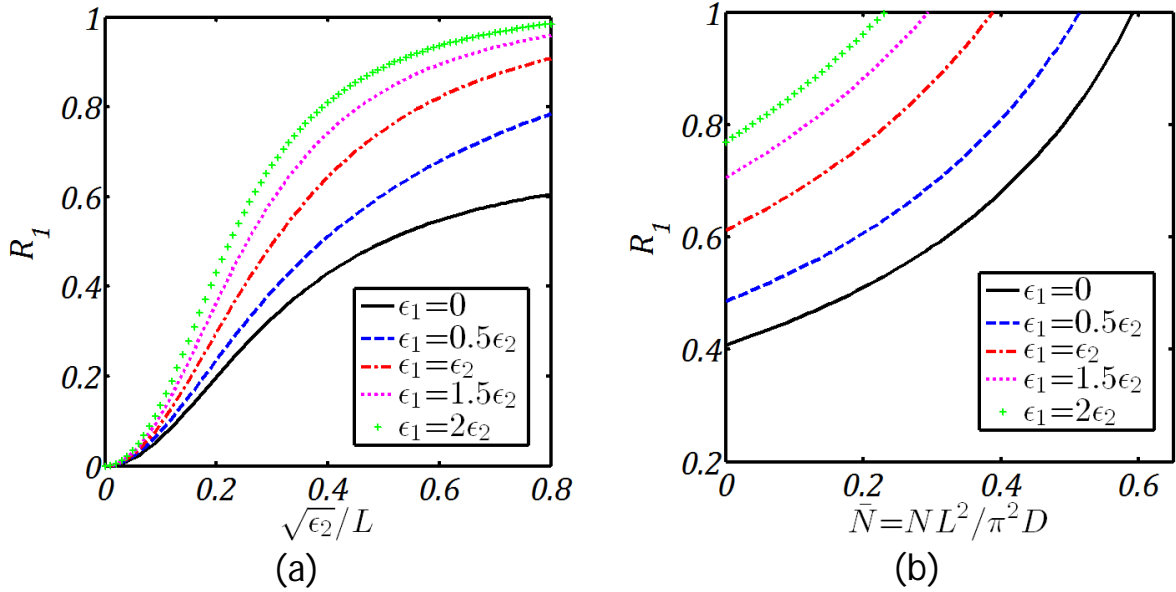

(a)  (b)

Figure 8: The correction factors for the prebuckling-fundamental natural frequency, $R_1$, of simple supported nonlocal Euler-Bernoulli beams as functions of (a) the nondimensional nonlocal parameter, $\sqrt{\epsilon_2}/L$, ($\lambda = \mu$, *i.e.* $\nu = 0.25$, and $\bar{N} = 0.05$) and (b) the nondimensional axial compressive load, $\bar{N} = NL^2/\pi^2D$, ($\lambda = \mu$, *i.e.* $\nu = 0.25$, and $\sqrt{\epsilon_2}/L = 0.4$).

Utilizing the obtained correction factors, the nondimensional prebuckling-fundamental natural frequency, $\bar{\omega}_1 = \omega_1\sqrt{mL^4/D}$, of simple supported Euler-Bernoulli beams as a function of the nondimensional axial compressive load, $\bar{N} = NL^2/\pi^2D$, is plotted when $\epsilon_1 = \epsilon_2$ and $\epsilon_1 = 2\epsilon_2$ in Figures 9(a) and 9(b), respectively. Moreover, the results of the local beam, $\sqrt{\epsilon_2}/L = 0$, are plotted in parallel to the results obtained by Shaat [29] where an excellent match can be observed. As shown in Figure 9, the nondimensional prebuckling fundamental natural frequency decreases with the increase in the nondimensional axial load value. Furthermore, at the critical buckling load value, *i.e.* when $N = N_{cr}$, the prebuckling fundamental natural frequency decreases to zero. Moreover, it can observed that the increase in the nondimensional nonlocal parameter is accompanied with a decrease in, both, the nondimensional prebuckling fundamental natural frequency and the critical buckling load. By comparing Figure 9(a) with Figure 9(b), the nondimensional prebuckling-fundamental natural frequency and the nondimensional critical buckling load decrease with the increase in the nonlocal parameters-ratio $\epsilon_1/\epsilon_2$.

The nondimensional natural frequencies of the first and second modes of the prebuckling vibration of simple supported Euler-Bernoulli beams under axial compressive loads are presented in Figure 10 as functions of the nondimensional axial load, $\bar{N} = NL^2/\pi^2D$, for different nonlocal parameters-ratio, $\epsilon_1/\epsilon_2$. It is clear that the increase in the nondimensional axial load results in reductions in the nondimensional prebuckling natural frequencies. Moreover, the nondimensional prebuckling natural frequencies decrease with the increase in the nonlocal parameters-ratio $\epsilon_1/\epsilon_2$. This can be attributed to the fact that the considered nonlocal parameters affect the beam with a softening mechanism. Moreover, it can be observed that the initiation of buckling is associated with a zero prebuckling fundamental natural frequency.

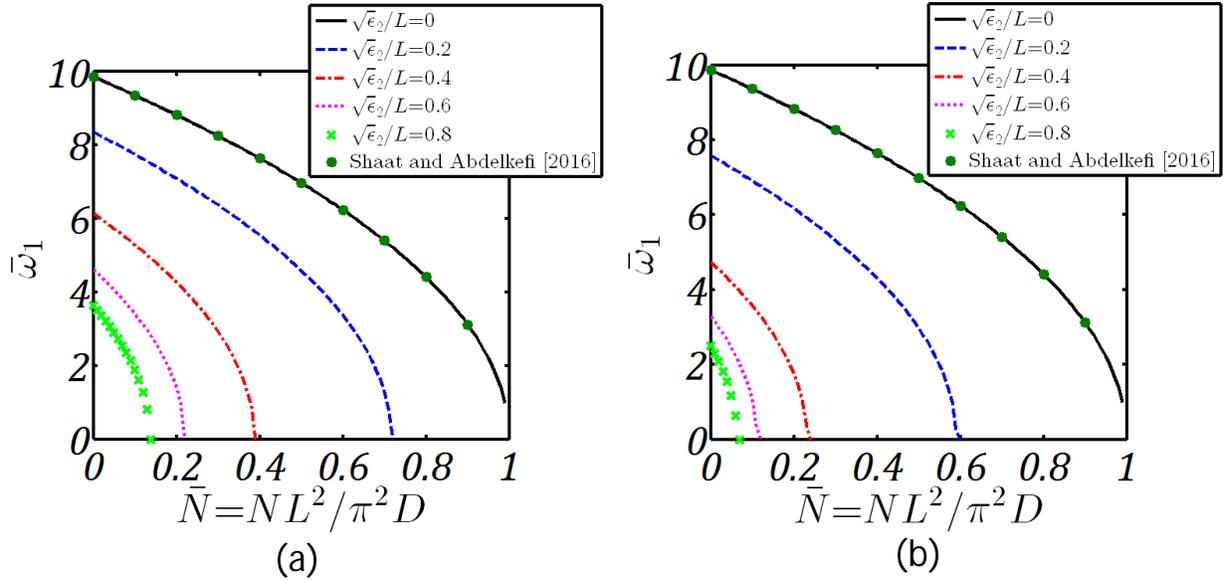

(a)                                                              (b)

Figure 9: The nondimensional prebuckling-fundamental natural frequency, $\bar{\omega}_1 = \omega_1\sqrt{mL^4/D}$, of simple supported Euler-Bernoulli beams as a function of the nondimensional axial compressive load, $\bar{N} = NL^2/\pi^2 D$, when (a) $\lambda = \mu$, *i.e.* $\nu = 0.25$, and $\epsilon_1 = \epsilon_2$ and (b) $\lambda = \mu$, *i.e.* $\nu = 0.25$, and $\epsilon_1 = 2\epsilon_2$.

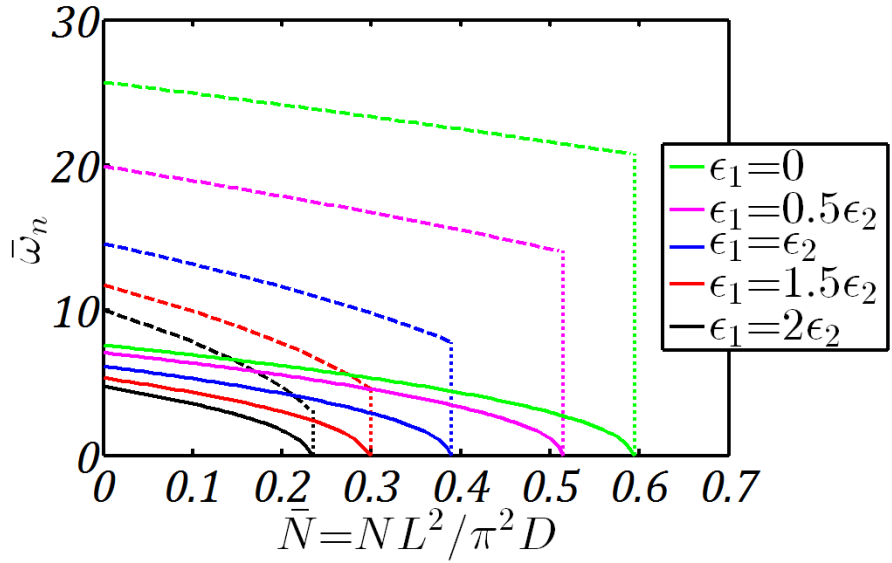

Figure 10: The nondimensional natural frequencies, $\bar{\omega}_n = \omega_n\sqrt{mL^4/D}$, of prebuckled Euler-Bernoulli beams under axial compressive loads as functions of the nondimensional applied axial load, $\bar{N} = NL^2/\pi^2 D$, ($\lambda = \mu$, *i.e.* $\nu = 0.25$, and $\sqrt{\epsilon_2}/L = 0.4$). *Solid lines denote the fundamental frequencies while dashed lines refer to the second mode frequencies.*

### 3.5 Buckling configurations

For simple supported nonlocal Euler-Bernoulli beams, the buckling configurations are obtained utilizing the correction factors of the different buckling modes as $R_n = \psi_n$, as shown to Table 2. To demonstrate that the effectiveness of the proposed approach to determine the buckling configurations of nonlocal Euler-Bernoulli beams, the buckling configurations when $\epsilon_1 = \epsilon_2$ are obtained compared to the results obtained by Emam [30], as presented in Figure 11. The exact same buckling configurations for the different nonlocal parameters as those presented in Emam [30] are obtained. The impacts of the nonlocal residuals on the buckling configuration of simple supported Euler-Bernoulli beams are depicted in Figure 12. Moreover, the merits of the general nonlocal theory over Eringen's nonlocal theory can be read from Figure 12. It is clear that the increase in the nonlocal parameters-ratio is accompanied with a decrease in the nondimensional beam deflection and its nondimensional critical buckling load.

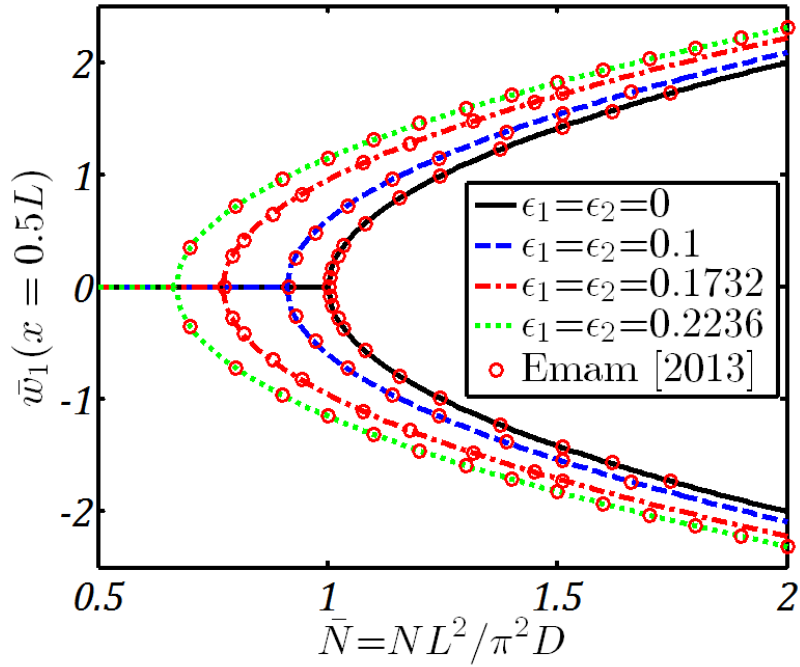

Figure 11: Model validation: the nondimensional fundamental buckling configuration, $\bar{w}_1(L/2) = w_1(L/2)\sqrt{A/I}$, of simple supported Euler-Bernoulli beams as a function of the nondimensional axial compressive load, $\bar{N} = NL^2/\pi^2 D$.

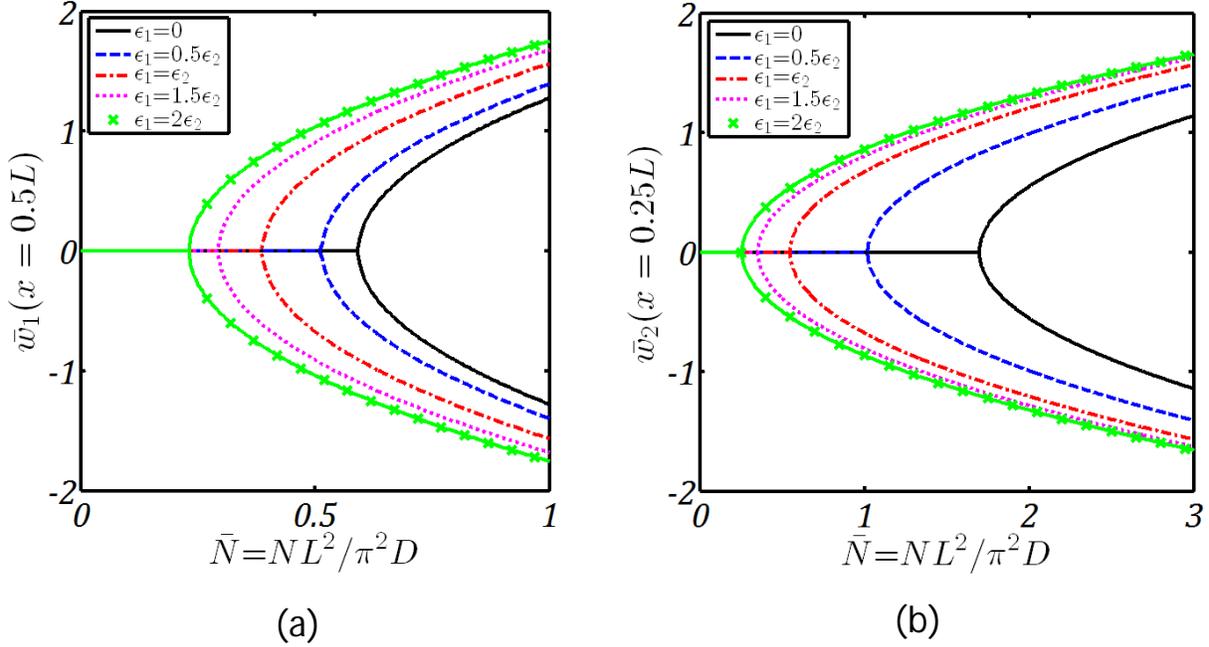

(a)                                      (b)

Figure 12: The nondimensional buckling configurations, $\bar{w}_n(x) = w_n(x)\sqrt{A/I}$, of simple supported Euler-Bernoulli beams as functions of the nondimensional axial compressive load, $\bar{N} = NL^2/\pi^2 D$, ($\lambda = \mu$, *i.e.* $\nu = 0.25$, and $\sqrt{\epsilon_2}/L = 0.4$).

## Conclusions

In this study, a new methodology that depends on the iterative-nonlocal residual approach was proposed to easily obtain the elastic nonlocal fields from their local counterparts without solving the nonlocal field equation. This methodology can resolve the existing complications of deriving solutions of the nonlocal field equation. Thus, instead of solving the field equation in the nonlocal field, the local fields were corrected for the nonlocal residuals.

In this study, correction factors for the local characteristics of Euler-Bernoulli beams including bending, vibration, and buckling characteristics were derived. These correction factors were utilized to directly obtain the nonlocal characteristics of Euler-Bernoulli beams without solving the beam's equation of motion. The correction factors were formed based on the general nonlocal theory which considers two distinct nonlocal parameters for isotropic-linear elastic continua. It was demonstrated that the general nonlocal theory outperforms Eringen's nonlocal theory in accounting for the impact of the beam's Poisson's ratio on its nonlocal bending, vibration, and buckling characteristics.

A parametric study on the impacts of the nonlocal parameters and the beam's Poisson's ratio on the bending, vibration, and buckling characteristics of nonlocal Euler-Bernoulli beams was performed. The obtained results revealed that, depending on the nonlocal parameters, Eringen's nonlocal theory may result in an inappropriate modeling of the nonlocal effects on Euler-Bernoulli beams. The general nonlocal theory, however, perfectly model the nonlocal fields in Euler-Bernoulli beams.